%% file: Ising2D_regular_v1.tex
\documentclass[aps,twocolumn,pre,showpacs,superscriptaddress]{revtex4-2}
\usepackage{bm}
\usepackage{graphicx}
\usepackage{amsmath}
\usepackage{amssymb}
\usepackage{color}
\usepackage{wasysym}
\usepackage{enumerate}
\usepackage[colorlinks]{hyperref}
\usepackage{ifthen}
\usepackage{xspace}
\usepackage[caption=false]{subfig}

\def\today{May 28, 2021; revised September 17, 2021}
\begin{document}

\title{Slow growth of magnetic domains helps fast evolution routes for 
out-of-equilibrium dynamics}

\author{Isidoro Gonz\'alez-Adalid Pemart\'{\i}n}\affiliation{Departamento de
  F\'\i{}sica Te\'orica, Universidad Complutense, 28040 Madrid, Spain}

\author{Emanuel Momp\'o}\affiliation{Departamento de Matem\'aticas, 
  Universidad Carlos III de Madrid, 28911 Legan\'es, Spain}

\author{ Antonio Lasanta}\affiliation{Departamento de \'Algebra. Facultad 
de Educaci\'on, Econom\'ia y Tecnolog\'ia de Ceuta, Universidad de 
Granada, Cortadura del Valle, s/n, 51001 Ceuta, Spain}%
  \affiliation{Grupo de Teor\'{\i}as de Campos 
  y F\'{\i}sica Estad\'{\i}stica, Instituto Gregorio Mill\'an, 
  Universidad Carlos III de Madrid, Unidad Asociada al Instituto de 
  Estructura de la Materia, CSIC, Spain}  

\author{Victor Mart\'in-Mayor}\affiliation{Departamento de F\'\i{}sica
  Te\'orica, Universidad Complutense, 28040 Madrid,
  Spain}\affiliation{Instituto de Biocomputaci\'on y F\'{\i}sica de Sistemas
  Complejos (BIFI), 50018 Zaragoza, Spain}

\author{Jes\'us Salas}\affiliation{Departamento de Matem\'aticas, 
  Universidad Carlos III de Madrid, 28911 Legan\'es, Spain}%
  \affiliation{Grupo de Teor\'{\i}as de Campos 
  y F\'{\i}sica Estad\'{\i}stica, Instituto Gregorio Mill\'an, 
  Universidad Carlos III de Madrid, Unidad Asociada al Instituto de 
  Estructura de la Materia, CSIC, Spain}  

\date{\today}

\begin{abstract}
Cooling and heating faster a system is a crucial problem in science,
technology and industry. Indeed, choosing the best thermal protocol to reach a
desired temperature or energy is not a trivial task. Noticeably, we find that
the phase transitions may speed up thermalization in systems where there are
no conserved quantities. In particular, we show that the slow growth of
magnetic domains shortens the overall time that the system takes to reach a
final desired state. To prove that statement, we use intensive numerical
simulations of a prototypical many-body system, namely, the 
two-dimensional Ising model.
\end{abstract}

\maketitle

%
%
\section{Introduction}

Nonequilibrium relaxation processes have been extensively studied during the
last decades. One important scope in this research is shortening the duration
of the heating (or cooling) transient process that precedes
thermalization. Indeed, annealing techniques are a popular tool to accelerate
the evolution towards the equilibrium---or stationary---state through a slow
temperature decrease (not only in physics: the famous simulated annealing
algorithm mimics in a computer the annealing strategy followed in the
laboratory~\cite{kirkpatrick:83}).  Thus, it is not surprising that recent
extensions of the counterintuitive Mpemba Effect~\cite{MO69}, allowing to cool
down faster the hotter of two systems (or heat up faster the cooler
system), have stirred considerable attention. Indeed, we now understand which
are the general conditions allowing for faster coolings, or faster heatings,
in Markovian systems~\cite{LR17,KRHM19}, granular matter~\cite{LVPS17,Tal19},
spin glasses~\cite{Bal19}, water~\cite{GLH19}, the quantum Ising spin model
\cite{NF19} and very recently the generalization to Markovian open quantum
systems~\cite{CLL21}. On these grounds, Amit and Raz have designed a novel
strategy, useful in systems with time-scale separation, in which precooling
the system results in a faster heating~\cite{GR20}.

Yet, time scale separation is not possible at a second-order phase transition,
where critical slowing-down~\cite{parisi:88,zinn-justin:05} evinces a
continuum of time scales [see, e.g., Eq.~\eqref{eq:exponential} below]. Under
these circumstances, unraveling the mechanism that drives the dynamics is the
key to potentially control the evolution.  The growth of the ordered domains
when the system enters the symmetry-broken phase~\cite{bray:94,ABCS07,Bi87}
emerges as a natural candidate for this mechanism. Energy is stored in ordered
domains whose size and growth arrest or accelerates the dynamics. In
particular, the relevance of the domain growth for the dynamic slowdown in
spin glasses is now clear~\cite{marinari:96,BB02,janus:08b,janus:16,%
janus:17b,janus:18,zhai-janus:20a,paga:21}.

Here we show that the overall relaxation time can be shortened in the absence
of time-scale separation by manipulating the system's internal structure of
ordered domains. We focus on the study of the ferromagnetic 
two-dimensional (2D) Ising spin
model.  In particular, we study through numerical simulations an unexplored
out-of-equilibrium 
heating protocol, in which the bath temperature starts below
the critical temperature and is later heated above the critical point.
Surprisingly, we find that this manipulation of the bath temperature induces a
speed-up in the energy evolution of the system, which is due to a slow
domain-growth process. The mechanism is illustrated by comparing different
initial preparation of the system.

This paper is organized as follows: In Sec.~\ref{sec:model} we 
recall some essential facts about the 2D Ising model and the 
quantities of interest. The one- and two-step thermal protocols are 
described in Sec.~\ref{sec:thermal}, where we also show the number of 
simulations performed for each protocol. In 
Sec.~\ref{sec:iso_evolution}, we discuss the isothermal evolution of a 
system in both the ferromagnetic and paramagnetic phases. 
Section~\ref{sec:leading} is devoted to show how the leading time corrections 
can be canceled out in the two-step protocol. In 
Sec.~\ref{sec:equilibration}, we discuss the speed-up of the 
equilibration procedure for the two-step protocol. 
The conclusions are contained in Sec.~\ref{sec:conclusions}. The most 
technical parts of our work are explained in two Appendixes: In 
Appendix~\ref{Sec:Algorithm}, we show how we have implemented the Monte
Carlo (MC) dynamics, and in Appendix~\ref{Sec:integrals}, we explain how 
to perform the spatial integrals of the two-point correlation function.  

%
%
\section{Model and quantities of interest} \label{sec:model} 

We consider the ferromagnetic Ising model in two dimensions, one of the most
thoroughly studied models in statistical
mechanics~\cite{mccoy:73}. Specifically, the spins $\sigma_{\bm{x}}=\pm 1$
occupy the nodes $\bm{x}$ of a square lattice of size $L\times L$ with
periodic boundary conditions. In our Hamiltonian, the spins interact with
their lattice nearest neighbors as
\begin{equation}\label{eq:H}
\mathcal{H}\;=\; -J\sum_{\langle \bm{x},\bm{y} \rangle} \sigma_{\bm{x}}
\sigma_{\bm{y}} \,.
\end{equation}
The thermal bath is described through the (dimensionless) inverse-temperature
$\kappa=J/(K_\mathrm{B} T)$. We have set $J=1$ energy units.
A second-order phase transition at
$\kappa_\mathrm{c}= \log(1 + \sqrt{2})/2$ separates the paramagnetic phase at
$\kappa< \kappa_\mathrm{c}$, from the ferromagnetic phase at
$\kappa>\kappa_\mathrm{c}$. 

We simulate two dynamic rules for the model in Eq.~\eqref{eq:H}: the
Metropolis and heat bath (HB) algorithms (see,
e.g.,~\cite{sokal:97,landau:05}).  Both belong to the so-called \emph{model A}
dynamic universality class~\cite{hohenberg:77}, where conserved quantities are
lacking. We choose $L=4096$ in our simulations~\footnote{We implement a
  multisite multispin coding strategy~\cite{fernandez:15}, adapted to the
  square lattice~\cite{fernandez:19} and to the comparatively high
  temperatures of this work (see Appendix~\ref{Sec:Algorithm}
  and~\cite{gonzalezadalid:22}).}, large enough to represent the thermodynamic
limit.  Our time step corresponds to a full-lattice sweep. 
The technical details about how we have implemented both dynamics 
are contained in Appendix~\ref{Sec:Algorithm}.

Special attention will be payed to the time evolution of the coherence length
$\xi(t)$, namely, the typical linear size of the ferromagnetic domains at time
$t$.
Note that  the correlation length $\xi_{\text{corr}}$ indicates
the spatial range of correlations \emph{within} a domain. Only in the
paramagnetic phase $\xi=\xi_{\text{corr}}$. The classification of length
scales in the ferromagnetic phase would be even subtler in the presence of
Goldstone bosons~\cite{josephson:66}.

Our second important quantity is the
energy density $E(t)$, a thermometric quantity~\cite{Bal19} for which the
equilibrium value $E_\text{eq}\equiv E(t\to\infty)$ is given by the well-known
Onsager result. Both $E(t)$ and $\xi(t)$ are obtained from the correlation
function
\begin{equation}\label{eq:C}
C(\bm{r};t) \;=\; \frac{1}{L^2}\, \sum_{\bm{x}}
\left\langle\, \sigma_{\bm{x}}(t)\sigma_{\bm{x}+\bm{r}}(t)\, \right\rangle\,, 
\end{equation}
where $\langle \ldots\rangle$ indicates the average over $N_\mathrm{R}$
independent trajectories or replicas obtained with 
the same thermal protocol (see next section for the number of 
replicas used in practice). 

Indeed, $E(t)= -2 C(\bm{r}_\circ;t)$, where
$\bm{r}_\circ=(0,1)$ [or $(1,0)$, because these 
are the two vectors spanning the square 
lattice], and $\xi(t)$ is computed from space integrals of
$C(\bm{r};t)$ \footnote{At $L=\infty$ and for 
  large distances $r$, $C(\bm{r};t)\sim {\mathcal G}(x,y)/r^\zeta $, where
  $x=r/\xi(t)$, $y=\xi(t)/\xi(t\to \infty)$ (for $\kappa \geq
  \kappa _\text{c}$, $y=0$), $\zeta$ is an exponent, and
  ${\mathcal G}(x,y)$ is a scaling function that decays
  super-exponentially in $x$ (exponentially, if $y=1$)~\cite{fernandez:18b}. 
  The pairs $\{{\mathcal G}(x,y),\ \zeta\}$ are different
  for $\kappa<\kappa_\text{c}$, $\kappa=\kappa_\text{c}$, and
  $\kappa>\kappa_\text{c}$.  We circumvent the parametrization of
  ${\mathcal G}$ by computing space integrals of $C(\bm{r};t)$ 
  \cite{janus:08b} (see Appendix~\ref{Sec:integrals})} (see also 
  Appendix~\ref{Sec:integrals}).

%
%
\section{Thermal protocols} \label{sec:thermal}  

\begin{figure}[t]
\centering\includegraphics[width=0.48 \textwidth]{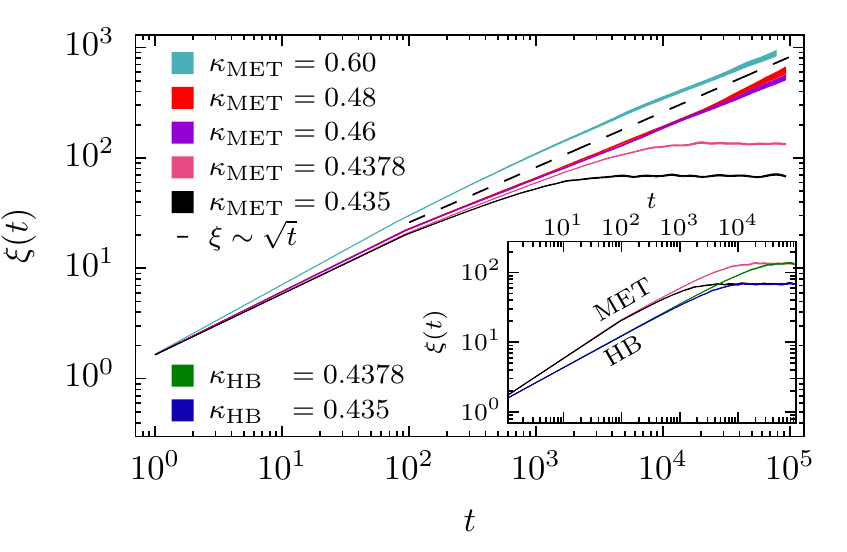}
\caption{\label{fig:xi} Coherence length $\xi$ as a function of time $t$, 
  as computed with the Metropolis algorithm for our one-step  
  protocol and several values of 
  $\kappa$ (the width of the curves is twice the statistical error; $\kappa$
  increases from bottom to top). Only in the paramagnetic phase,
  $\kappa<\kappa_{\mathrm{c}}\approx 0.44068679$, the coherence length reaches
  its equilibrium value at long times. In the ferromagnetic phase
  $\xi(t)\sim \sqrt{t}$ at long times (dashed line). 
  Inset: Comparison of the HB
  and Metropolis dynamics in the paramagnetic phase. Although $\xi(t)$ grows
  significantly faster for Metropolis, the equilibrium limit at long times is
  the same for both dynamics.} 
\end{figure}

We consider two distinct thermal protocols, and each of them is 
simulated twice (with the Metropolis and HB dynamics). 
In our \emph{one-step} 
protocol, a fully disordered spin configuration 
(corresponding to infinite temperature)
is put in contact with a thermal bath at
inverse-temperature $\kappa$, at the initial time $t=0$. The coherence length
$\xi(t)$ grows with time (see Fig.~\ref{fig:xi}) until (in the paramagnetic
phase) its $t\to\infty$ limit is approached.

%
%

In the \emph{two-step} protocol, a fully disordered spin configuration is
initially placed at an inverse temperature in the ferromagnetic phase,
$\kappa_{\text{start}}>\kappa_{\text{c}}$, where it evolves until $\xi(t)$
reaches a target value $\xi_{\text{start}}$. At that point, which corresponds
with our initial time $t=0$, the bath temperature is instantaneously raised to
enter the paramagnetic phase, $\kappa_{\text{end}}<\kappa_{\text{c}}$, and
kept fixed afterwards.  Note that $\xi_{\text{start}}$ may be larger than
$\xi_{\text{end}} \equiv\xi(t\to\infty;\kappa_{\text{end}})$, and that a 
one-step protocol with $\kappa<\kappa_\text{c}$ is a particular 
case of the two-step protocol with $\xi_{\text{start}}=0$ and
$\kappa_{\text{end}}=\kappa$.

\begin{figure}[t]
  \includegraphics[width=0.48 \textwidth]{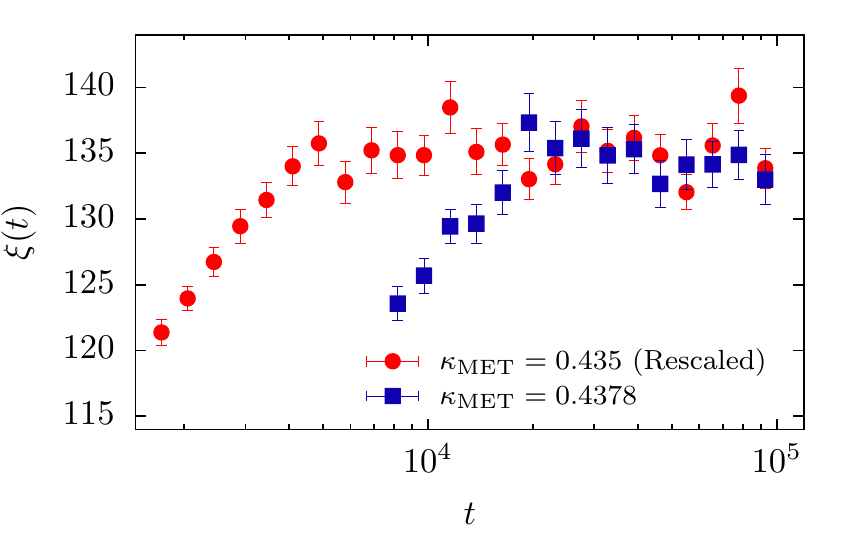}
  \caption{\label{fig:factor_escala} 
    Time evolution of the coherence length in
    the paramagnetic phase for the one-step 
    protocol for the Metropolis
    dynamics. The data for $\kappa=0.435$ are multiplied by the scale factor
    $(\kappa_{\text{c}}-0.435)/(\kappa_{\text{c}}-0.4378)$. Indeed, after the
    rescaling, the $\kappa=0.435$ data agree at long times with the
    $\kappa=0.4378$ data, showing that we are in the scaling region (indeed,
    the thermal critical exponent is $\nu=1$). For later reference, let us
    quote the value $\xi_\text{end}=135(2)$ for the long-time limit at
    $\kappa=0.4378$.
    }
\end{figure}

We take $\kappa_{\text{end}}=0.435$ and $0.4378$, 
where $\xi_{\text{end}}$ is very large (see Fig.~\ref{fig:xi}). Indeed, the
product $(\kappa_{\mathrm{c}}-\kappa_{\text{end}})\,
\xi_{\text{end}}(\kappa_{\text{end}})$ remains constant as
$\kappa_{\text{end}}\to\kappa_{\mathrm{c}}$~\cite{mccoy:73}. 
This behavior is explained by the fact that the correlation length 
coincides with the coherence length $\xi$ in the paramagnetic phase, 
and the value $\nu=1$ of the thermal critical exponent
(remember that $\xi_\text{end}$ is the coherence length at
$\kappa_\text{end}$ for long times, when the equilibrium is reached). Of
course, this scale invariance is found only close enough to the 
critical point (in the so-called scaling region). 
Figure~\ref{fig:factor_escala} shows that we are indeed working in 
the scaling region.

For each one of our thermal protocols, and our
two dynamics (Metropolis and HB), we have simulated 
$N_\mathrm{R}=256$ replicas with a few exceptions:

\begin{enumerate}
\item The one-step protocol with $\kappa=0.435$ (resp.\/ $\kappa=0.4378$),
where we have made $1536$ (resp.\/ $4096$) replicas.
\item The two-step protocol with $\kappa_{\text{start}}=0.46$,
$\kappa_{\text{end}}=0.4378$ and $\xi_{\text{start}}=120$, where we simulate
1024 replicas.  
\end{enumerate}

We have chosen to simulate a larger number of replicas in the
paramagnetic phase in order to reduce the statistical error in the correlation
function $C(\bm{r};t)$. This error reduction is fundamental for computing the
spacial integrals of $C(\bm{r};t)$, from which we compute the 
coherence length (see Appendix~\ref{Sec:integrals}).

%
%

%
%
\section{The isothermal evolution} \label{sec:iso_evolution}

As shown in Fig.~\ref{fig:xi}, the dynamic behavior in the two phases 
is very different. In the paramagnetic phase, $\kappa<\kappa_\text{c}$, both
$\xi(t)$ and $E(t)$ approach exponentially their equilibrium
value~\cite{parisi:88,fernandez:19}:
\begin{equation}\label{eq:exponential}
O(t;\kappa)\;=\; O_\infty(\kappa)\,\left[\,1\ -\ 
\int_1^{\tau_{\kappa}}\rho_O(\tau,\kappa)\, \text{e}^{-t/\tau}\,
\mathrm{d}\tau \right]\,,
\end{equation}
where $O(t;\kappa)$ stands for $E(t;\kappa)$ or $\xi(t;\kappa)$, and
$\rho_O(\tau,\kappa)$ is a continuous distribution of autocorrelation times
$\tau$. 
The largest timescale $\tau_\kappa$ (which diverges at $\kappa_{\text{c}}$)
ensures that finite-time corrections decay as $\text{e}^{-t/\tau_{\kappa}}$, 
or faster.
Although using Metropolis or HB does make a difference, see
the inset of Fig.~\ref{fig:xi}, the equilibrium value at large 
$t$ is independent of the dynamics.

%
%
\begin{figure}[htb]
  \centering
  \subfloat{
    \centering
    \includegraphics[width=0.48 \textwidth]{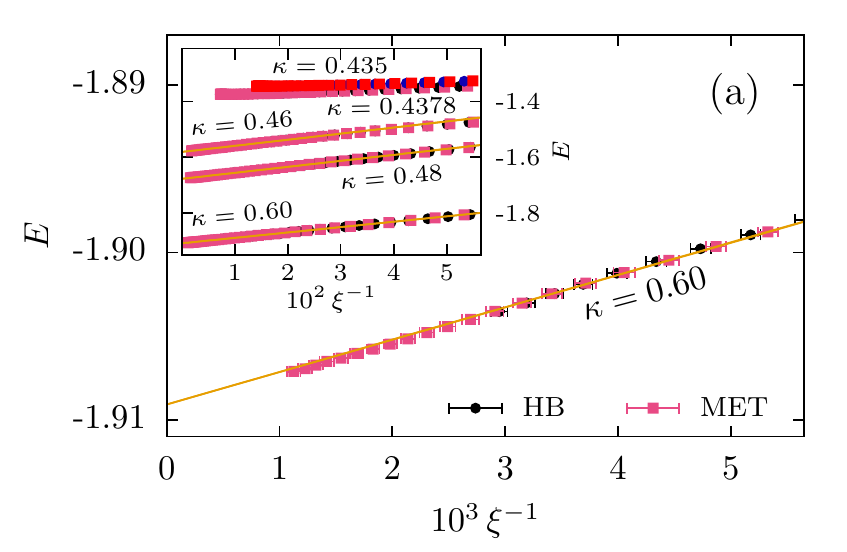}
  }\\
  \vspace{-5mm}
  \subfloat{
    \centering
    \includegraphics[width=0.48 \textwidth]{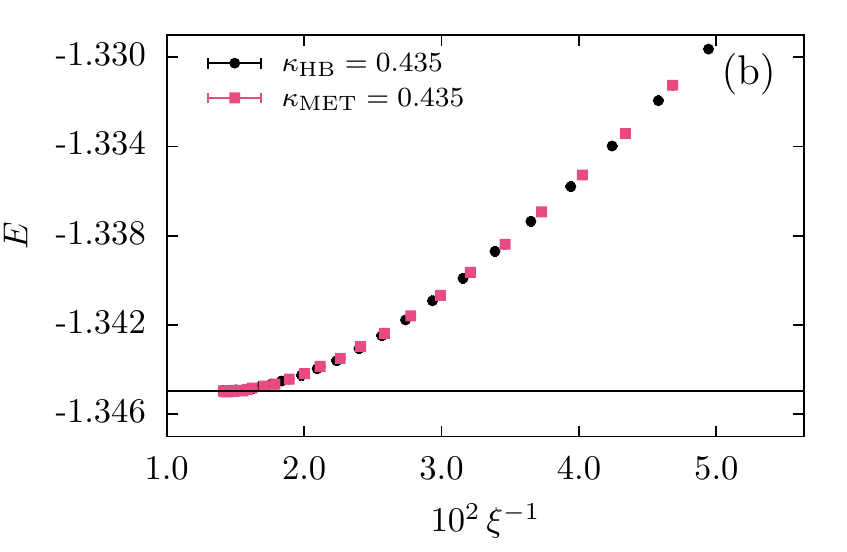}
  }
  \caption{\label{fig:energy-xi} 
    Energy density $E(t)$ from our one-step 
    protocol as a function of $1/\xi(t)$. 
    (a) Data for $\kappa=0.6$ and both Metropolis and HB dynamics. 
     Both data sets fall on a single curve. The continuous line is a fit to the 
     {\em Ansatz} $E(t)=E_{\text{eq}}+b/\xi(t)$. The only free parameter is 
     the slope $b$, and we include only data with $\xi(t)>200$.   
     Inset: Data for several values of $\kappa$, whose value increases
     from top to bottom. Notice that for a given value of $\kappa$, 
     the Metropolis and HB data sets fall on the same curve. 
     The continuous lines for values of $\kappa$ in the ferromagnetic 
     phase are fits to the {\em Ansatz} 
     $E(t;\kappa)=E_{\text{eq}}(\kappa)+b_\kappa/\xi(t;\kappa)$ for $\kappa$. 
     Again, the only free parameters are the slopes $b_\kappa$, and in each 
     fit, we include only data with $\xi(t)>20$. 
     (b) Data for $\kappa=0.435$ (paramagnetic phase). The solid 
     horizontal line is the exact Onsager solution. The error bars are smaller
     than the symbols. Notice the null slope at 
     equilibrium, which implies that the growth of the magnetic domains 
     does not affect the energy.
  }
\end{figure}

Instead, in the ferromagnetic phase, the largest timescale exists only as a
finite-size effect. Indeed, when $\kappa>\kappa_\text{c}$ and barring
short-time corrections, domains grow as $\xi\sim\sqrt{t}$~\cite{bray:94} until
$\xi\sim L$ (see Fig.~\ref{fig:xi}).

Now, it is well known that in the ferromagnetic phase, and excluding fast
initial relaxations, $E(t)$ and $\xi(t)$ are tightly
connected~\cite{parisi:88}: $(E(t)-E_{\text{eq}})\propto 1/\xi(t)$, 
because the excess energy is located at the boundaries of the magnetic 
domains (and moreover, the lower critical dimension is $1$ for Ising systems). 
This is the behavior found for $\kappa > \kappa_\text{c}$: see 
Fig.~\ref{fig:energy-xi}(a), and the lower three curves in the inset of
that figure. 
Perhaps more surprisingly, we find that this strong connection extends 
to the paramagnetic phase where, close to the equilibrium, 
the  magnetic domains grow without changing the system energy 
[see the top two curves displayed in the inset of 
Fig.~\ref{fig:energy-xi}(a)].
Indeed [see Fig.~\ref{fig:energy-xi}(b)], if one represents $E(t)$ as a
function of $1/\xi(t)$ for $\kappa < \kappa_\text{c}$, the data from 
Metropolis and HB dynamics fall on a single curve (yet, the two 
time behaviors are remarkably different; see the inset of Fig.~\ref{fig:xi}).
Furthermore, notice the null slope of such a curve at equilibrium endpoint [see
Fig.~\ref{fig:energy-xi}(b)], which implies that the growth of the 
magnetic domains does not affect the energy.

The connection between $E(t)$ and $\xi(t)$ suggests an intriguing
possibility. Given that $\xi(t)$ grows faster in the ferromagnetic phase (see
Fig.~\ref{fig:xi}), it is conceivable that $E(t)$ could equilibrate
\emph{faster} in the paramagnetic phase through an excursion to the
ferromagnetic phase.

%
%
\section{Canceling leading time corrections} \label{sec:leading}

%
%
\begin{figure}[t]
  \centering
  \subfloat{
    \centering
    \includegraphics[width=0.48 \textwidth]{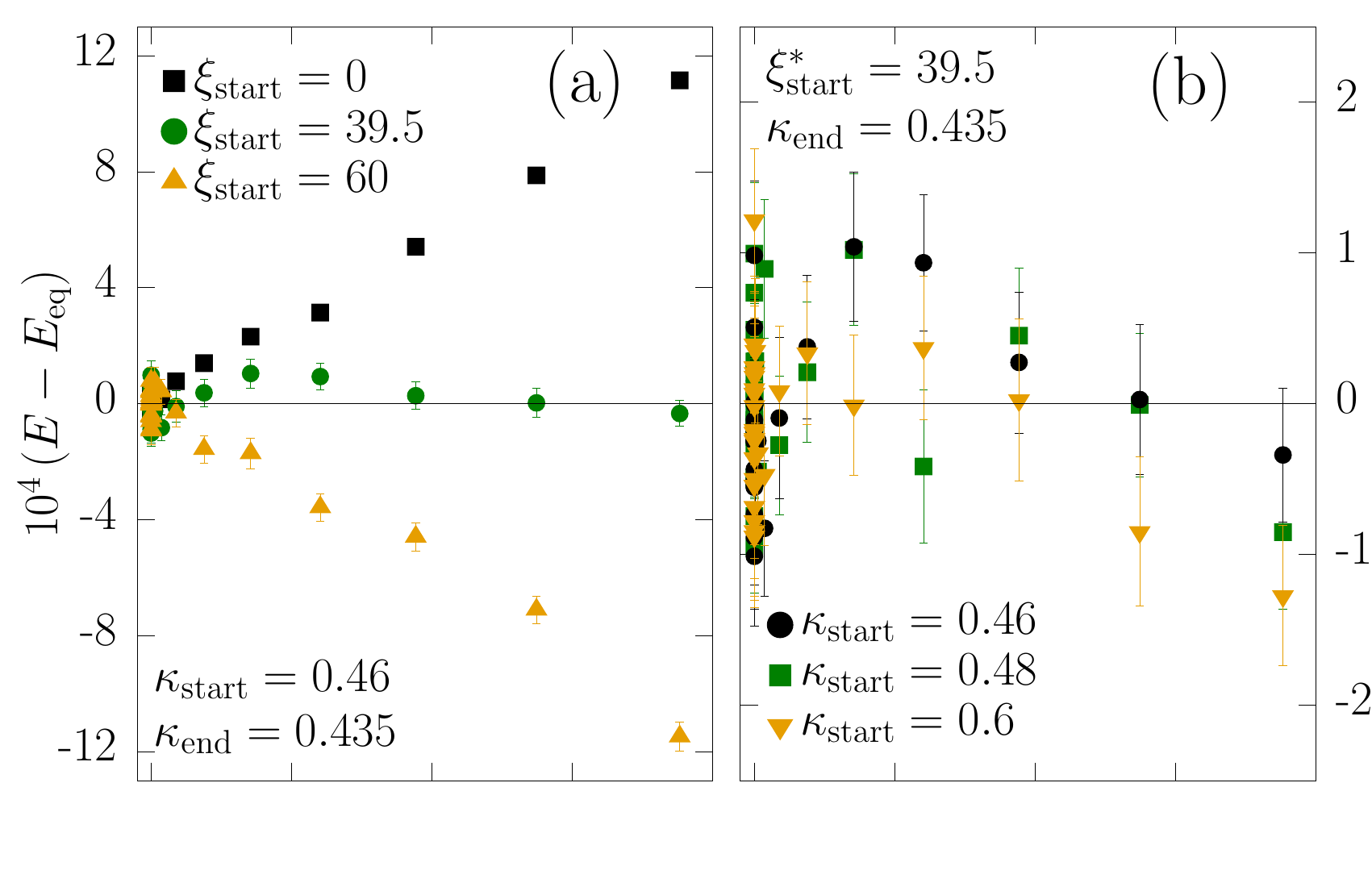}
  }\\
  \vspace{-1cm}
  \subfloat{
    \centering
    \includegraphics[width=0.48 \textwidth]{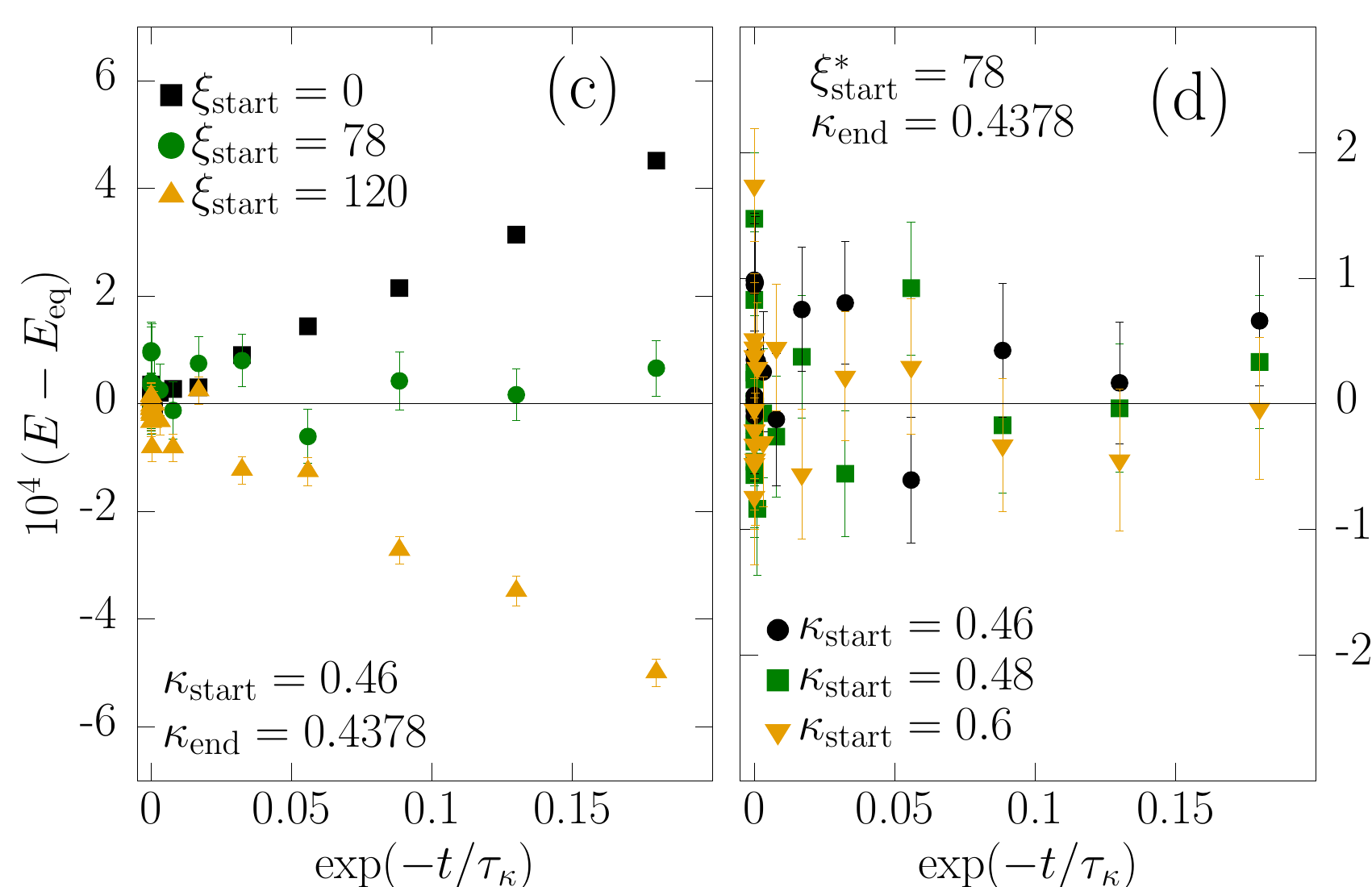}
  }
  \caption{\label{fig:punto-dulce} Excess energy $E(t)-E_{\text{eq}}$ vs
    $\mathrm{e}^{-t/\tau_\kappa}$, as obtained with Metropolis dynamics for
    our two-step protocol.
    (a) Data for
    $\kappa_{\text{start}}=0.46$ and $\kappa_{\text{end}}=0.435$, as computed
    for several values of $\xi_{\text{start}}$. 
    (b) Data for $\xi_{\text{start}}^*=39.5$ and
    $\kappa_{\text{end}}=0.435$, as computed for several values of
    $\kappa_{\text{start}}$. 
    (c) As in panel (a), for $\kappa_{\text{start}}=0.46$ and 
    $\kappa_{\text{end}}=0.4378$. 
    (d) As in panel (b), 
    for $\xi_{\text{start}}^*=78$ and $\kappa_{\text{end}}=0.4378$.  }
\end{figure}

\begin{figure}[t]
    \centering
  \subfloat{
    \centering
    \includegraphics[width=0.48 \textwidth]{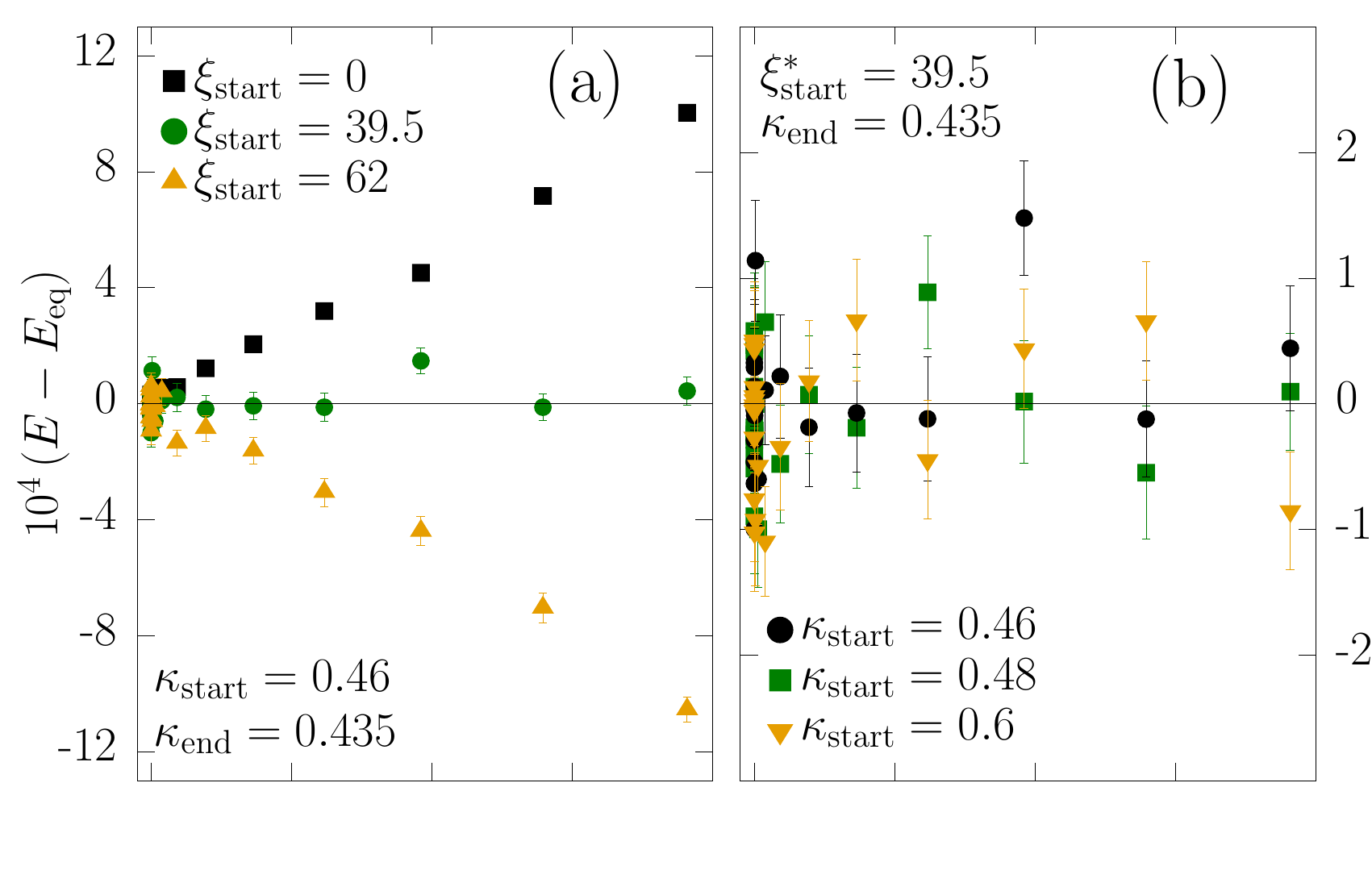}
  }\\
  \vspace{-1cm}
  \subfloat{
    \centering
    \includegraphics[width=0.48 \textwidth]{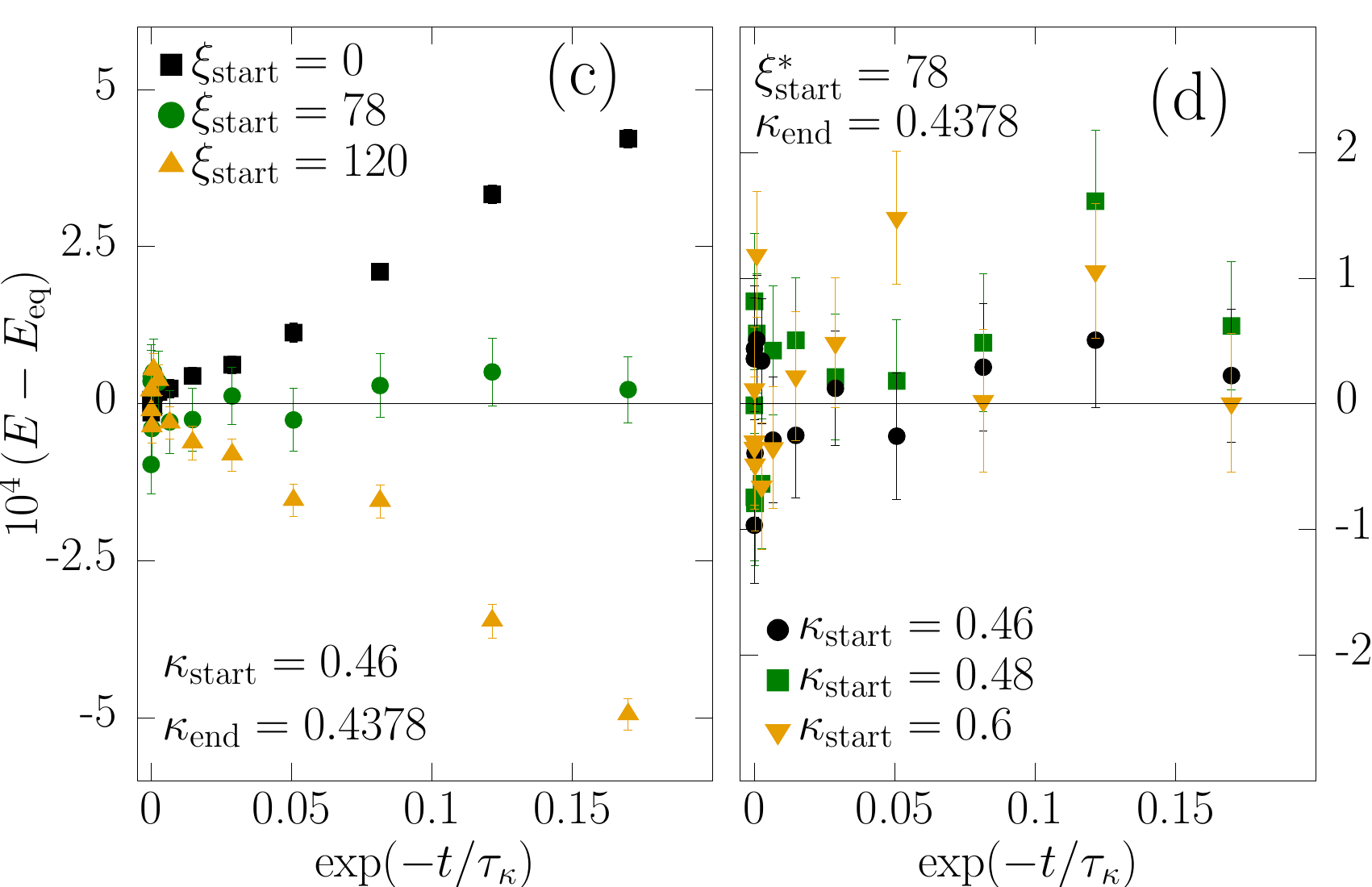}
  }
  \caption{\label{fig:punto-dulceSM} 
    Excess energy $E(t)-E_{\text{eq}}$ vs
    $\mathrm{e}^{-t/\tau_\kappa}$, as obtained with the HB dynamics
    for our two-step protocol. 
    (a) Data for
    $\kappa_{\text{start}}=0.46$ and $\kappa_{\text{end}}=0.435$, as computed
    for several values of $\xi_{\text{start}}$. (b) Data for
    $\xi_{\text{start}}^*=39.5$ and $\kappa_{\text{end}}=0.435$, as computed
    for several values of $\kappa_{\text{start}}$. 
    (c) As in panel (a), for $\kappa_{\text{start}}=0.46$ and
    $\kappa_{\text{end}}=0.4378$.  (d) As in panel
    (b), for $\xi_{\text{start}}^*=78$ and $\kappa_{\text{end}}=0.4378$.}
\end{figure}

As suggested by Eq.~\eqref{eq:exponential}, in the paramagnetic phase, the
two-step protocol behaves at long times as (see
Fig.~\ref{fig:punto-dulce} for the Metropolis dynamics and 
Fig.~\ref{fig:punto-dulceSM} for the HB one):
\begin{equation}\label{eq:amplitud}
  E(t)-E_{\text{eq}} \;=\; {\mathcal A}\, 
  (\xi_{\text{start}})\,\mathrm{e}^{-t/\tau_\kappa}\,,
\end{equation}
where ${\mathcal A}(\xi_{\text{start}})$ is an amplitude (the one-step
protocol is recovered for $\xi_{\text{start}}=0$).  Interestingly enough, 
${\mathcal A}(\xi_{\text{start}})$ changes sign as the initial 
coherence length grows. This phenomenon is clearly seen 
in Figs.~\ref{fig:punto-dulce}(a)-\ref{fig:punto-dulce}(c) 
and~\ref{fig:punto-dulceSM}(a)-\ref{fig:punto-dulceSM}(c).
It follows that there exists a $\xi_{\text{start}}^*$ such that 
${\mathcal A}(\xi_{\text{start}}^*)=0$, which entails an exponential 
speed-up. It turns out that $\xi_{\text{start}}^*$ is independent of 
$\kappa_{\text{start}}$ [see 
Figs.~\ref{fig:punto-dulce}(b)-\ref{fig:punto-dulce}(d) 
and~\ref{fig:punto-dulceSM}(b)-\ref{fig:punto-dulceSM}(d)].
However, computing $\xi_\text{start}^*$ is difficult
because of the statistical uncertainty in $\mathcal{A}(\xi_\text{start})$.

%
%

In order to estimate $\xi_\text{start}^*$ we have taken several values of
$\xi_\text{start}$ around our initial guess for $\xi_\text{start}^*$, obtained
from Fig.~\ref{fig:punto-dulce} (Metropolis) and 
Fig.~\ref{fig:punto-dulceSM} (HB).
The value of $\tau_\kappa$ in Eq.~\eqref{eq:amplitud} is estimated by fitting
to the {\em Ansatz} \eqref{eq:amplitud} our most accurate data, namely, that 
with $\xi_\text{start}=0$. We then estimate $\mathcal{A}(\xi_\text{start})$ 
using the same {\em Ansatz} \eqref{eq:amplitud} with $\tau_\kappa$ fixed to 
the previously obtained value. We have chosen the fitting window as 
$0\leq\text{e}^{-t/\tau_\kappa}\leq 0.2$.

Our strategy consists in finding an interval
$(\xi_\text{start}^+,\xi_\text{start}^-)$ where $\xi_\text{start}^*$ is
contained (see Tables~\ref{tabla:0435} and~\ref{tabla:04378}). We require
$\mathcal{A}$ to be positive (resp.\/ negative) with high 
probability (i.e., larger in absolute value than twice its statistical error) 
at $\xi_\text{start}^+$ (resp.\/ $\xi_\text{start}^-$).
Finally, we estimate the value of $\xi_{\text{start}}^*$
as the average of $\xi_\text{start}^+$ and $\xi_\text{start}^-$, and its error
as the semi-difference of $\xi_\text{start}^+$ and $\xi_\text{start}^-$ (see
Table~\ref{tabla:xi}).

%
%
\begin{table}[htb]
   \subfloat[Metropolis]{
     \begin{tabular}{c c c}
       \hline\hline
       $\xi_\text{start}$ &  $\mathcal{A}\ \times 10^4$ & 
       $|\mathcal{A}|/\Delta\mathcal{A}$ \\ \hline
       $37\phantom{.5}$ & $10.6(18)$           & $5.9$\\
       $39.5$           & $\phantom{-}0.8(20)$ & $0.4$\\
       $42\phantom{.5}$ & $-4.6(18)$           & $2.6$\\
       \hline\hline
   \end{tabular}}
   \hfill
   \subfloat[Heat bath]{
     \begin{tabular}{c c c}
       \hline\hline
       $\xi_\text{start}$ &  $\mathcal{A}\ \times 10^4$ & 
       $|\mathcal{A}|/\Delta\mathcal{A}$ \\ \hline
       $36\phantom{.5}$ & $\phantom{-}13(2)\phantom{.91}$ & $\phantom{1}6.5$\\
       $39.5$           & $\phantom{-2}3(2)\phantom{.91}$ & $\phantom{1}1.5$ \\
       $50\phantom{.5}$ & $-26.3(19)$                    & $13.8$\\
       \hline\hline
   \end{tabular}}
  \caption{\label{tabla:0435} 
    Values of $\mathcal{A}(\xi_\text{start})$
    obtained with $\kappa_\text{start}=0.46$ and $\kappa_{\text{end}}=0.435$
    for the Metropolis (a) and HB (b) dynamics.
    $\Delta\mathcal{A}$ is the
    statistical error obtained from the fit to Eq.~\eqref{eq:amplitud}. The
    value of $\tau_{\kappa}$ used to fit the data is 433(14) for Metropolis,
    and 1470(50) for HB.
    }
\end{table}

%
%
\begin{table}[htb]
  \subfloat[Metropolis]{
    \begin{tabular}{|c c c|}
      \hline\hline
      $\xi_\text{start}$ &  $\mathcal{A}\times 10^4$ & 
      $|\mathcal{A}|/\Delta\mathcal{A}$ \\ \hline
     $70$& $\phantom{-}5(2)\phantom{.91}$&$2.5$\\
     $79$&          $-2(2)\phantom{.91}$ &$1\phantom{.5}$\\
     $87$&       $-5.0(19)$              &$2.6$\\
      \hline
  \end{tabular}}
  \hfill
  \subfloat[Heat bath]{
    \begin{tabular}{|c c c|}
      \hline
      $\xi_\text{start}$ &  $\mathcal{A} \times 10^4$ & 
      $|\mathcal{A}|/\Delta\mathcal{A}$ \\ \hline
     $70\phantom{.5}$&$\phantom{-}9(2)$&$4.5$\\
     $81.5$          &$-1(3)$          &$0.3$\\
     $92\phantom{.5}$&$-8(3)$          &$2.7$\\
      \hline\hline
  \end{tabular}}
  \caption{\label{tabla:04378} Values of $\mathcal{A}(\xi_\text{start})$
    obtained with $\kappa_\text{start}=0.46$ and $\kappa_{\text{end}}=0.4378$
    for the Metropolis (a) and HB (b) dynamics. $\Delta\mathcal{A}$
    is the statistical error obtained from the fit to
    Eq.~\eqref{eq:amplitud}. The value of $\tau_{\kappa}$ used to fit the
    data is 2000(100) for Metropolis, and 6500(300) for HB.}
\end{table}

%
%
\begin{table}[htb]
  \begin{tabular}{|c  c | c  c|}
    \hline\hline
    \multicolumn{2}{|c|}{$\kappa_\text{end}=0.435$} & 
    \multicolumn{2}{|c|}{$\kappa_\text{end}=0.4378$} \\ \hline
        MET & HB & MET & HB \\ \hline
        40(3) & 43(7) & 79(9) & 81(11) \\ \hline\hline
  \end{tabular}
  \caption{\label{tabla:xi} $\xi_{\text{start}}^*$ estimate for { the}
    Metropolis (MET) and HB dynamics
    for each value of $\kappa_\text{end}$. The center
    value and its error are obtained from the semisum and semidifference of
    the extreme values from the
    Tables~\ref{tabla:0435} and~\ref{tabla:04378}.}
\end{table}

The final results for $\xi_\text{start}^*$ shows a compatibility between both
dynamics (see Table~\ref{tabla:xi}). The values of $\xi_\text{start}^*$ 
should be compared with the coherence length at equilibrium 
$\xi_\text{end}(\kappa)$,
namely, $\xi_\text{end}(0.435)=69(1)$ and $\xi_\text{end}(0.4378)=135(2)$,
both for the Metropolis dynamics.

Furthermore, when $\kappa_{\text{end}}$ is varied, both $\xi_{\text{start}}^*$
and the equilibrium coherence length $\xi_{\text{end}}$ scale as
$1/(\kappa_{\mathrm{c}}-\kappa_{\text{end}})$. 

These two traits, namely, independence
of $\kappa_{\text{start}}$ and the correct scaling dimension for
$\xi_{\text{start}}^*$, are hallmarks of universality. Indeed, we speculate
that the scale-invariant ratio that we find for the Metropolis dynamics
\begin{equation}
\lim_{\kappa\to\kappa_{\mathrm{c}}^-} 
\frac{\xi_{\text{start}}^*}{\xi_{\text{end}}} \;=\; 0.59(7)
\end{equation}
will be common to all models with scalar order parameter in the 
\emph{model A} dynamic universality
class~\cite{hohenberg:77}. Our results for the HB dynamics are 
indeed consistent with this speculation.

\section{Equilibration speed-up} \label{sec:equilibration}

\begin{figure}[htb]
  \centering\includegraphics[width=0.48 \textwidth]{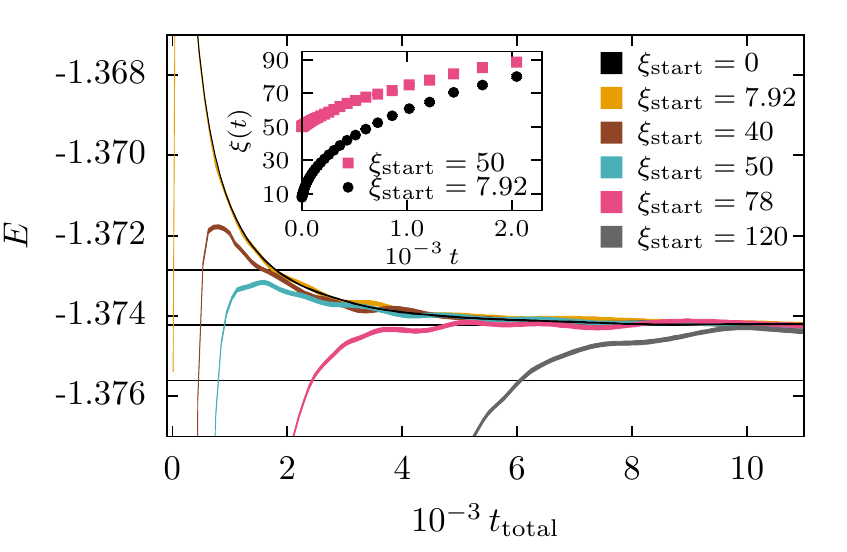}
  \caption{\label{fig:banda} Energy density as a function of the total time
    $t_{\text{total}}$ elapsed since the beginning of the 
    two-step protocol, as computed with the Metropolis dynamics
    for $\kappa_{\text{start}}=0.46$, $\kappa_{\text{end}}=0.4378$,
    and several values of $\xi_{\text{start}}$. These values of 
    $\xi_{\text{start}}$ increase from top to bottom.
    The width of the curves is twice the statistical errors.
    We show only the evolution during the second step of the protocol. 
    The three horizontal lines correspond to
    $E_{\text{eq}}$ multiplied by $1.001, 1$ and $0.999$. 
    Inset: Coherence length during the two-step protocol
    ($\kappa_{\text{start}}=0.46$, $\kappa_{\text{end}}=0.4378$) as a function
    of the time $t$ elapsed since the temperature changed. We show data for
    $\xi_{\text{start}}^{\text{Kovacs}}\approx 7.92$ (lower curve), as well as
    for $\xi_{\text{start}}=50$ (upper curve; this value yields the minimum
    equilibration time; see Fig.~\ref{fig:entrada}). }
\end{figure}

\begin{figure}[t]
  \centering\includegraphics[width=0.48 \textwidth]{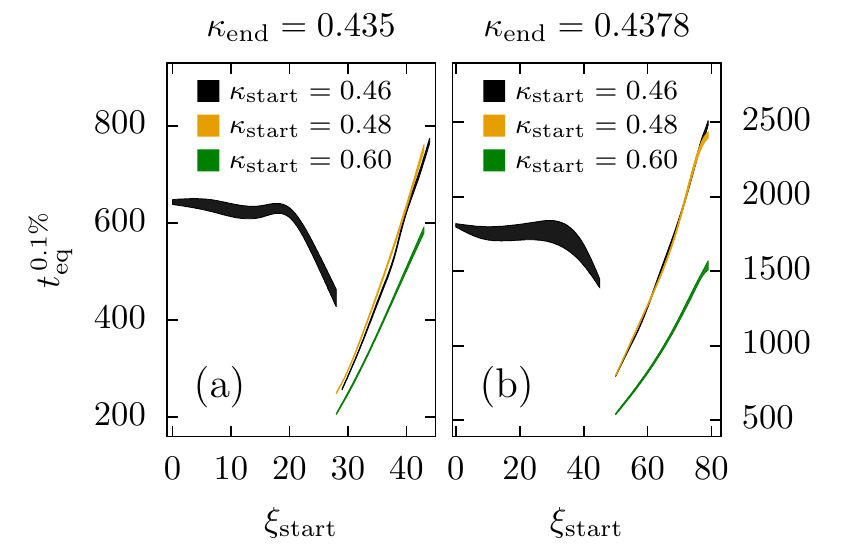}
  \caption{\label{fig:entrada} Equilibration time $t^{0.1\%}_{\text{eq}}$
    versus $\xi_{\text{start}}$ for the two-step protocol with
    $\kappa_{\text{end}}=0.435$ (a) and $\kappa_{\text{end}}=0.4378$
    (b). We show data for three values of $\kappa_{\text{start}}$
    and the Metropolis dynamics. The width of
    the curves is twice the statistical errors. The discontinuity is due to
    the nonmonotonic time behavior of $E(t_{\text{total}})$; see
    Fig.~\ref{fig:banda}. Note that, after the discontinuity, the curves for
    $\kappa_{\text{start}}=0.46$ and $\kappa_{\text{start}}=0.48$ are 
    very close.
    }
\end{figure}

Let us put to use the existence of an exponential speed-up. From now
on, we shall be referring to $t_{\text{total}}$, namely, the time spent
by the system at \emph{both} $\kappa_{\text{start}}$ and
$\kappa_{\text{end}}$.  In fact, we operationally define the
equilibration time $t^{0.1\%}_{\text{eq}}$ as the time such that
$E(t_{\text{total}})$ differs from $E_{\text{eq}}$ by less than 0.1\%
for any $t_{\text{total}}>t^{0.1\%}_{\text{eq}}$; see
Fig.~\ref{fig:banda}~\footnote{The discussion will clarify that the
  speed-up effect can be found as well for other values of the inverse
  temperature $\kappa_{\text{end}}$ and of the chosen width (of
  $0.1\%$) for the equilibration band around $E_{\text{eq}}$.
}. 
In other words, the equilibration time $t^{0.1\%}_{\text{eq}}$ 
corresponds to the \emph{last} time the energy crosses either the line
$1.001E_{\text{eq}}$ or the line $0.999E_{\text{eq}}$ 
(see Fig.\ref{fig:entrada}).
We need this operational definition because, strictly speaking, 
thermal equilibrium is unreachable in any finite time.

%
%

Now, let us consider the situation for the two-step protocol with
$\xi_{\text{start}} < \xi_{\text{start}}^*$ and where ${\cal
  A}(\xi_{\text{start}}) > 0$ [cf., Eq.~\eqref{eq:amplitud}]. In this range of
small $\xi_{\text{start}}$ we encounter
$\xi_{\text{start}}=\xi_{\text{start}}^{\text{Kovacs}}$ where the energy at
the time of the temperature change equals $ E_{\text{eq}}$ (this is the
appropriate situation to study the Kovacs effect~\cite{kovacs:79,BB02}). Now,
for those protocols with $\xi_{\text{start}}^{\text{Kovacs}}<
\xi_{\text{start}}< \xi_{\text{start}}^*$ , the energy at the time the
temperature changes is \emph{smaller} than $E_{\text{eq}}$~\footnote{Because
  the coherence length during the first step of the protocol gets larger than
  $\xi_{\text{start}}^{\text{Kovacs}}$, recall Figs.~\ref{fig:xi}
  and~\ref{fig:energy-xi}.} but, at long times, $E(t_{\text{total}})$ tends to
$ E_{\text{eq}}$ from \emph{above} because ${\cal A}(\xi_{\text{start}}) >
0$. It follows that $E(t_{\text{total}})$ cannot be monotonic. In fact (see
Fig.~\ref{fig:banda}), $E(t_{\text{total}})$ for those protocols first grow to
a local maximum, then decrease towards $E_{\text{eq}}$.

%
%

%
%
\begin{figure}[htb]
  \includegraphics[width=0.48 \textwidth]{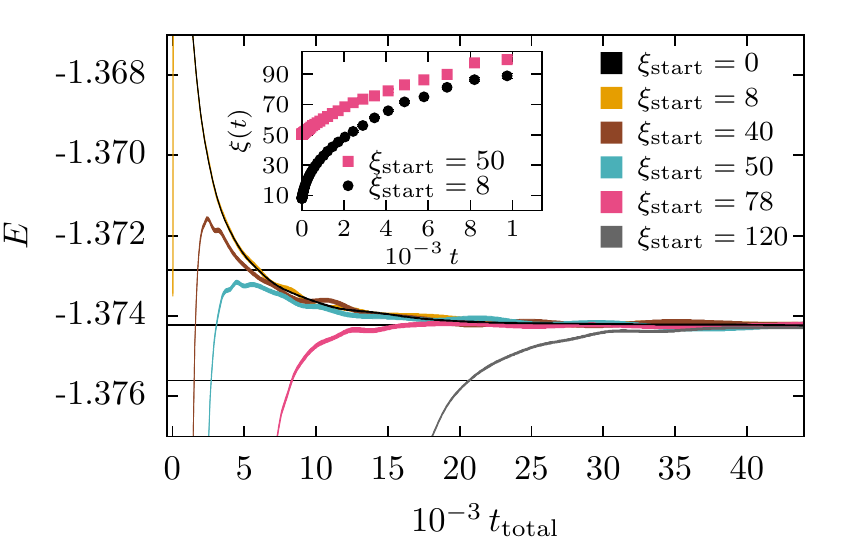}
  \caption{\label{fig:bandaSM} Energy density as a function of the total time
    $t_{\text{total}}$ elapsed since the beginning of the
    two-step protocol, as computed with the HB dynamics
    for $\kappa_{\text{start}}=0.46$, $\kappa_{\text{end}}=0.4378$,
    and several values of $\xi_{\text{start}}$. Plot definitions are as
    in Fig.~\ref{fig:banda}. 
    Inset: Coherence length during the two-step protocol
    ($\kappa_{\text{start}}=0.46$, $\kappa_{\text{end}}=0.4378$) as a function
    of the time $t$ elapsed since the temperature changed. We show data for
    $\xi_{\text{start}}^{\text{Kovacs}}\simeq 8$ (lower curve), as well as
    for $\xi_{\text{start}}=50$ (upper curve; this value yields the minimum
    equilibration time; see Fig.~\ref{fig:entradaSM}).}
\end{figure}

The local maximum of $E(t_{\text{total}})$ entails a discontinuity for the
equilibration time $t^{0.1\%}_{\text{eq}}$ (cf., Fig.~\ref{fig:entrada}) at
the value of $\xi_{\text{start}}$ for which the height of the local maximum of
$E(t_{\text{total}})$ coincides with the upper limit of the $0.1\%$ band
around $ E_\text{eq}$~\footnote{For smaller values of $\xi_{\text{start}}$,
  $E(t_{\text{total}})$ successively enters, exits and re-enters the
  band. According to our operational definition, $t^{0.1\%}_{\text{eq}}$ is
  the time of the last entrance in the band.}.  Therefore, the equilibration
time $t^{0.1\%}_{\text{eq}}$ for the two-step protocols just after the
discontinuity could be shorter than its one-step 
counterpart by a factor of three.

\begin{figure}[t]
  \includegraphics[width=0.48 \textwidth]{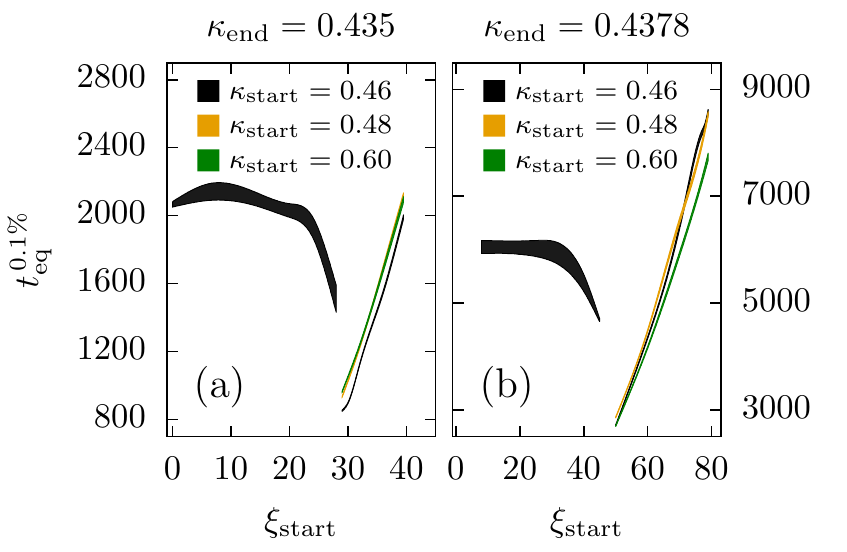}
  \caption{\label{fig:entradaSM} Equilibration time $t_{\text{eq}}^{0.1\%}$ as
    function of $\xi_{\text{start}}$ for the two-steps protocol with
    $\kappa_{\text{end}}=0.435$ (a) and
    $\kappa_{\text{end}}=0.4378$ (b). We show data for three
    values of $\kappa_{\text{start}}$ and the HB dynamics.  
    Plot definitions are as in Fig.~\ref{fig:entrada}. 
    The discontinuity is due to the nonmonotonic time behavior of
    $E(t_{\text{total}})$, see Fig.~\ref{fig:bandaSM}.
    Note that, after the discontinuity, the curves for
    $\kappa_{\text{start}}=0.48$ and $\kappa_{\text{start}}=0.60$ are
    very close in (a), and the curves for $\kappa_{\text{start}}=0.46$
    and $\kappa_{\text{start}}=0.48$ are very close in (b).}
\end{figure}

We can repeat the above computations for the HB dynamics. If we
represent the energy density vs the total simulation time $t_\text{total}$,
we obtain Fig.~\ref{fig:bandaSM}. The plot of the equilibration time 
$t_{\text{eq}}^{0.1\%}$ as a function of $\xi_{\text{start}}$ is given by 
Fig.~\ref{fig:entradaSM}. The behavior of these two quantities is analogous to
the one found for Metropolis. The main difference is the time scale. Indeed, 
Metropolis approaches equilibrium faster than HB.

%
%

%
%
\section{Conclusions} \label{sec:conclusions}

We have shown that precooling may result in faster equilibration at higher
temperatures in a prototypical system without timescale separation (namely,
the ferromagnetic 2D Ising model at its critical point). The
driving mechanism is the tight connection between the internal energy and the
size of the ferromagnetic domains, characteristic of a broken-symmetry
phase~\cite{parisi:88}. Indeed, the size of the domains can be manipulated to
our advantage through a nonequilibrium protocol. In particular, we have found
an exponential speed-up in the equilibration of the energy. The speed-up
arises when the size of the magnetic domains formed during the excursion to
the low-temperature phase is a well-defined fraction of the \emph{equilibrium}
correlation-length at the higher temperature. We have found numerical evidence
for the universality of this fraction (presumably, the universality is
restricted to \emph{model A} dynamics with scalar order parameter
\cite{hohenberg:77}). Namely, we have
shown how to design a practical protocol that exploits this universal
mechanism.

\begin{acknowledgments}
  This work was partially supported by Ministerio de Econom\'ia, Industria y
  Competitividad (MINECO, Spain), Agencia Estatal de Investigaci\'on (AEI,
  Spain), and Fondo Europeo de Desarrollo Regional (FEDER, EU) through Grants
  No.  PGC2018-094684-B-C21, No.  FIS2017-84440-C2-2-P and No.
  MTM2017-84446-C2-2-R. 
  A.L. and J.S. were also partially supported by Grant No.
  PID2020-116567GB-C22 AEI/10.13039/501100011033.
  A.L. was also partly supported by Grant No~A-FQM-644-UGR20 Programa operativo 
  FEDER Andaluc\'{\i}a 2014--2020. 
  J.S. was also partly supported by the Madrid Government
  (Comunidad de Madrid-Spain) under the Multiannual Agreement with UC3M in the
  line of Excellence of University Professors (EPUC3M23), and in the context
  of the V~PRICIT (Regional Programme of Research and Technological
  Innovation). I.G.-A.P. was supported by the Ministerio de Ciencia, 
  Innovaci\'on y
  Universidades (MCIU, Spain) through FPU Grant No. FPU18/02665.
\end{acknowledgments}

\appendix
%
%
\section{Implementation of multisite multispin algorithm}
\label{Sec:Algorithm}

Nowadays, many CPUs support 256-bit words in their streaming extensions. This
allows us to perform basic Boolean operations (AND, XOR, etc.) in parallel for
all the 256 bits, in a single clock cycle. We can take advantage from this
parallelization by codifying 256 distinct spins in the same 256-bit
word. This strategy is called multispin
coding~\cite{jacobs:81}. Furthermore, we simulate a \emph{single}
system, so we pack 256 spins from the same lattice. This is known as
MUlti-SIte (MUSI) multispin coding (or Synchronous multispin
coding~\cite{newman:99}). The main problem with MUSI multispin coding is
generating 256 independent random numbers for the 256 spins coded in a word. A
careless approach to the generation of these 256 random numbers would break
the parallelism, thus making MUSI multispin coding useless.

For the sake of clarity, we explain first our geometrical set up, and then
describe how we solved the problem of the independent random number for each
spin in the 256-bit word.

\subsection{Our multispin coding geometry}\label{SubSec:MUSI}

In our MUSI simulation, we packed 256 spins, from the same lattice, in a
single 256-bits word, forming a superspin~\cite{fernandez:15}. Specifically,
we employed the packing introduced in Ref. \cite{fernandez:19}. We present the
packing for a $M^2$-bit computer word (we shall specialize to $M=16$, and
$M^2=256$ bits). This packing transforms the original square lattice of size
$L\times L$ into a superspin lattice of size $(L/M)\times(L/M)$. The physical
coordinates $\bm{x}=(x,y)$ and the superspin coordinates $(i_x,i_y)$ are
related by:
\begin{equation}\label{eq:reticulo}
  \begin{split}
  x \;=\;\ & b_x\frac{L}{M}+i_x,\text{ where}\\
  & b_x\;=\;0,1,\cdots,M-1; \text{ and}\\
  & i_x\;=\;0,1,\cdots, \frac{L}{M}-1\, .
  \end{split}
\end{equation}
\begin{equation}\label{eq:reticulo2}
  \begin{split}
  y \;=\;\ & b_y\frac{L}{16}+i_y,\text{ where}\\
  & b_y\;=\;0,1,\cdots,M-1; \text{ and}\\
  & i_y\;=\;0,1,\cdots, \frac{L}{M}-1\, .
  \end{split}
\end{equation}

%
%
\begin{figure}[t]
 \includegraphics[width=0.48 \textwidth]{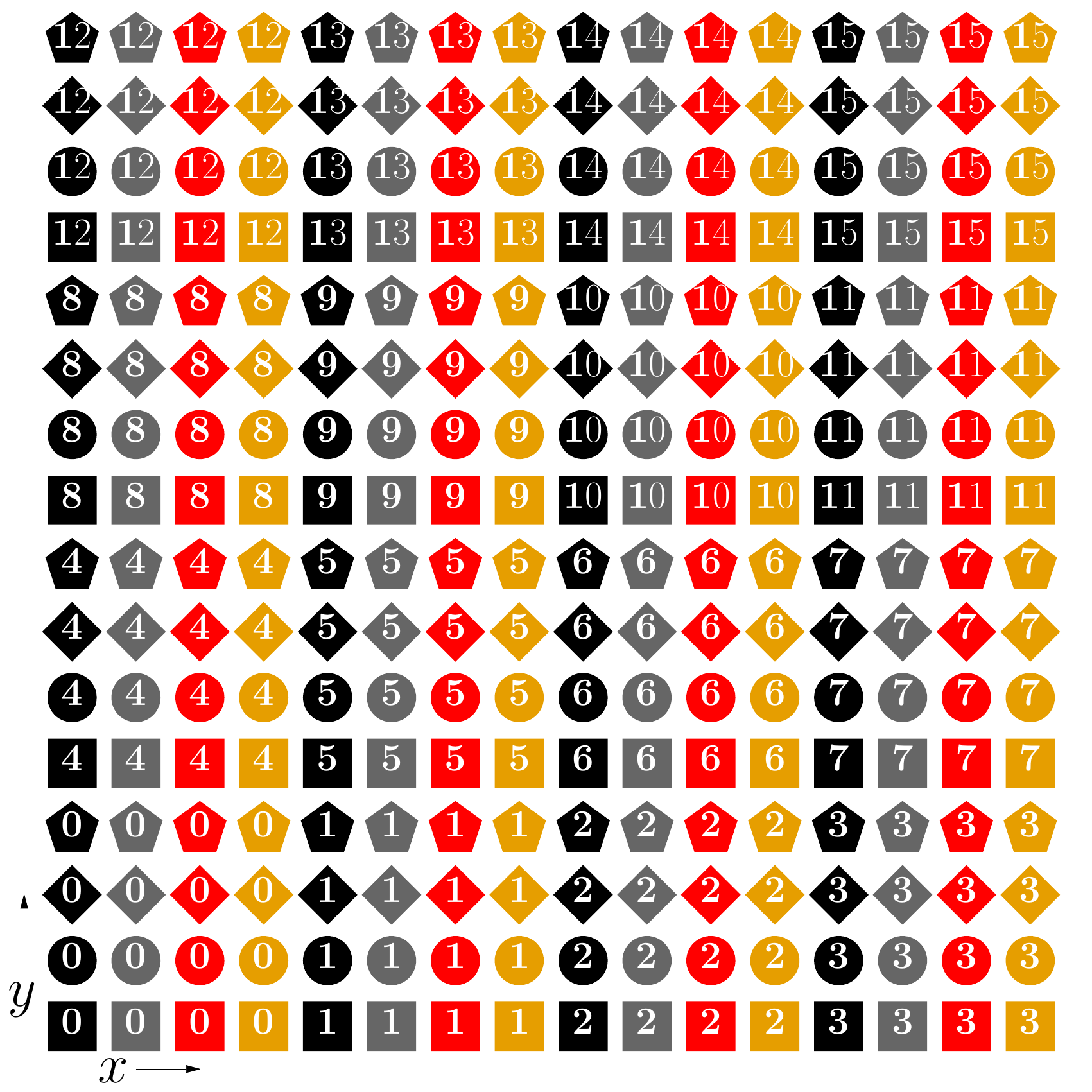}
  \caption{\label{fig:packing} Graphical representation of the packing process
    of a square lattice of size $16\times 16$ using 16-bit computer word
    (hence, $M=4$ in this example). The packing results into a $16/4\times
    16/4$ superspin lattice. The symbols represent the $i_x$ coordinate, and
    the colors represent the $i_y$ coordinate. Therefore, all spins with same
    symbol and color are codified in the same superspin. The number inside the
    symbol is the bit index $i_b=Mb_y+b_x$. Consider the spin at site $(x,y)$
    and its neighbor along the (say) forward $x$-direction, with $x$
    coordinate $(x+1) \mod L$. This neighbor is packed in the forward
    $x$-direction neighbor superspin, with $(i_x+1)\mod (L/M)$ coordinate. If
    $i_x<L/M-1$, the coordinates $i_y,b_x$ and $b_y$ remain unchanged [if
      $i_x=L/M-1$, $b_x$ changes when going forward along the $x$-direction as
      $b_x\rightarrow (b_x+1) \mod M$].}
\end{figure}

As a consequence, $M^2$ physical coordinates $(x,y)$ are assigned to the very
same superspin coordinates $(i_x,i_y)$. The bit index $i_b = Mb_y + b_x$ is
the position inside the superspin (so $0\leq i_b<M^2$), and unambiguously
identifies the physical coordinates (see Fig.~\ref{fig:packing} for a
graphical example of the packing process of a $L=16$ square lattice using
$M=4$).

There is a crucial feature of our chosen geometry. Take a superspin
$(i_x,i_y)$. Each of the $M^2$ spins packed in the superspin $(i_x,i_y)$ has a
nearest neighbor along the (say) forward $x$-direction in the original
lattice. All these $M^2$ neighboring spins are themselves packed in the same
superspin, namely, the nearest neighbor in the forward $x$-direction in the
superspin lattice (see Fig.~\ref{fig:packing}).

Another important property of Eqs.~\eqref{eq:reticulo}
and~\eqref{eq:reticulo2} is that the parity of the superspin site $i_x+i_y$
and the original site parity $x+y$ coincide (provided that $L/M=2n$, with
$n\in\mathbb{N}$). Note that the square lattice is bipartite: with our
nearest-neighbor interactions, sites of even parity interact only with
odd-parity sites (and vice versa). This feature suggests a checkerboard
updating scheme, in which all even sites are update in the first step and odd
sites are updated afterwards. Indeed, keeping the (say) odd spins fixed, the
ordering of the update of the even spins is immaterial. In particular, we may
update simultaneously the 256 spin that we pack in a superspin.

We define our time unit as a full-lattice sweep, in which the even sublattice
is updated first and the odd superspins are updated afterwards.

\subsection{Random numbers}\label{SubSec:random}

For reasons of clarity, we shall be referring always to a single bit,
$s_{\bm{x}}(t)$, in the superspin [$s_{\bm{x}}(t)= 1$ if the spin it codes is
  1 at time $t$, while the bit is zero if the spin is $-1$].  All the Boolean
operations that we shall explain below are performed simultaneously on the 256
bits $s_{\bm{x}}(t)$ contained in the superspin.

The scope of the game is obtaining a change bit $B_{\bm{x}}(t)$, which is one
if (and only if) the spin at site $\bm{x}$ is to be flipped at MC 
time $t$:
\begin{equation}
s_{\bm{x}}(t+1)\;=\,s_{\bm{x}}(t)\text{ XOR } B_{\bm{x}}(t)\, .
\end{equation}

The computation of $B_{\bm{x}}(t)$ is naturally decomposed in two steps (the
first is a deterministic step, while the second step involves randomness):
\begin{enumerate}
\item In the first step, we compute the energy change $\Delta E$ that flipping
  spin $s_{\bm{x}}(t)$ would cause.  Now, $\Delta E$ can take five values,
  $\Delta E \in \lbrace -8, -4, 0, 4, 8\rbrace$. Hence using standard Boolean
  operations, one computes five bits $\lbrace
  c_{\bm{x}}^{-8}(t),\ c_{\bm{x}}^{-4}(t),\ c_{\bm{x}}^0(t),\ c_{\bm{x}}^4(t),\ c_{\bm{x}}^8(t)\rbrace$
  in such a way that only the bit with superscript equal to $\Delta E$ is
  one. The remaining four bits are, of course, zero.
\item The second step depends on the dynamics, either Metropolis or HB,
  and determines whether or not the spin-flip is allo\-wed. This is
  represented by other five bits $\lbrace
  b_{\bm{x}}^{-8}(t),\ b_{\bm{x}}^{-4}(t),\ b_{\bm{x}}^0(t),\ b_{\bm{x}}^4(t),\ b_{\bm{x}}^8(t)
  \rbrace$, which are set to one with probability $p(\Delta E)$, as we explain
  below.
\end{enumerate}

We obtain $B_{\bm{x}}(t)$ from these bits as (for brevity, we will omit
their dependence on $\bm{x}$ and $t$):
\begin{equation}\label{eq:B}
  \begin{split}
    B \;=\; & \ [c^8 \text{ AND } b^8 ] \text{ OR }  [c^4 \text{ AND } b^4] \\
    & \text{ OR } [c^0 \text{ AND } b^0] \text{ OR } [c^{-4} \text{ AND }
      b^{-4}] \\
    & \text{ OR } [c^{-8} \text{ AND } b^{-8}].
  \end{split}
\end{equation}

\subsubsection{The probabilities}\label{SubSubSec:update_spin}

The probabilities for setting to one the random bits $b$ depend on the precise
algorithm we are using:
\begin{enumerate}
\item[a)] Metropolis: \begin{equation}\label{eq:p(MET)} p_\text{MET}(\Delta
  E)\;=\;\text{min}\lbrace 1,e^{-\Delta E/T}\rbrace\, ,
  \end{equation}
\item[b)] Heat-bath: \begin{equation}\label{eq:p(HB)} p_\text{HB}(\Delta
  E)\;=\;\displaystyle\frac{e^{-\Delta E/T}}{1+e^{-\Delta E/T}}\, .
\end{equation}
\end{enumerate}

\subsubsection{Some remarks on statistical independence}
\label{SubSubSec:bits_update}

Equation~\eqref{eq:B} tells us that we are going to use 
only one of the five bits
$b$, although we cannot know in advance which one. Indeed, the $\text{AND}$
conditions make irrelevant those bits $b$ whose superscript differ from
$\Delta E$. Hence, the five $b$ bits do not need to be mutually independent
(instead, the $b$ bits corresponding to different $\bm{x}$ or $t$ must be
statistically independent). We can make use of this observation to reduce the
number of $b$ bits that we need to compute:
\begin{enumerate}
\item[a)] In the Metropolis dynamics, we only need two bits because $
  p_\text{MET}(\Delta E)=1$ for $ \Delta E \leq 0$.  Therefore, we set
  $b^{-8}=b^{-4}=b^0=1$.
\item[b)] In the HB dynamics, because $p_\text{HB}(\Delta
  E)=1-p_\text{HB}(-\Delta E)$, we set $b^{-8}=\text{ NOT } b^{8}$ and
  $b^{-4}=\text{ NOT } b^{4}$.
\end{enumerate}

\subsubsection{Generating independent random bits with arbitrary probability}
\label{SubSubSec:generating_bits}

As we have seen, we need to generate a stream of independent random bits,
which are set to one with a probabi\-lity $p$ given by Eq.~\eqref{eq:p(MET)}
or~\eqref{eq:p(HB)}. The textbook solution~\cite{newman:99} fails the
independence requirement (unless $p$ can be exactly represented with a small
number of bits). On the other hand, the rather high critical temperature in
our problem makes the Gillespie method~\cite{fernandez:15} inefficient because
$p$ is too large. We solve this problem by adapting a
strategy~\cite{gonzalezadalid:22} that somewhat interpolates between the
textbook and the Gillespie methods.

To simplify the explanation, let us consider only one of the five bits $\lbrace
b^{-8},\ b^{-4},\ b^0,\ b^4,\ b^8 \rbrace$, for example $b^4$. We obtain $b^4$
as $b^4=d_1 \text{ OR } d_2$, where $d_1$ and $d_2$ are two independent random
bits which are set to one with pro\-babilities $p_1$ and $p_2$,
respectively. It is easy to check that the probability $ p(4)$ for having
$b^4=1$ is $ p(4)=p_1+p_2(1-p_1)$.  Hence, if we choose $p_1$ in some
convenient way (see below), we need to set $p_2$ as
\begin{equation}\label{eq:p2}
  p_2 \; =\; \frac{{ p(4)}-p_1}{1-p_1}\, .
\end{equation}
The overall idea is the following: if we can efficiently ge\-nerate $d_1$ with
a probability $p_1$ very close to (but smaller than) $ p(4)$, we will find
ourselves with a $p_2$ small enough for an efficient use of the Gillespie
method.

Specifically, we require that $p_1$ be exactly representable with $m$ bits
\begin{equation}
  \begin{split}
    p_1 & \;=\; \frac{k^*+1}{2^m}, \text{ with } k^* \in\mathbb{N},\ 
0\;\leq\; k^*<2 ^{m},\ \text{ and } \\
    & \frac{k^*+1}{2^m}\;\leq\; { p(4)} \;<\; \frac{k^*+2}{2^m}\, .
  \end{split}
\end{equation}
We obtain $d_1$ by generating an integer-valued random number $k$, $0\leq k<
2^m$, with uniform probability. We set $d_1=1$ if $k\leq k^*$.

Notice that $m$ determines the efficiency of the algorithm. On one hand,
enlarging $m$ can be detrimental because we generate $k$ by obtaining $m$
independent random bits which are set to one with $50\%$ probability. On the
other hand, a large $m$ allows us to have $p_1$ very close to $p(\Delta E)$,
and hence a very small $p_2$. A tradeoff needs to be found, by minimizing the
total number of calls to our random number generator.

An important simplification is that we are allowed to use the same random
integer $k$ for all the $\Delta E$, only the threshold $k^*(\Delta E)$
varies. In this way, we obtain all the bits $d_1$ for every $b^8$ and $b^4$.

As for the second bit $d_2$, we have two different probabilities $p_2$, one
for $b^8$ { [$p_2(8)$],} and other for $b^4$ { [$p_2(4)$]}. Let
$p_\text{max}=\text{max}\lbrace p_2(4), p_2(8)\rbrace$ and
$p_\text{min}=\text{min}\lbrace p_2(4), p_2(8)\rbrace$, so we can implement
the Gillespie method for $p_\text{max}$, which gives the bits $d_2$ for the
$\Delta E$ with bigger $p_2$ probability. The bit $d_2$ corres\-ponding to
$p_\text{min}$ is set to one if two conditions are met: (1) the bit $d_2$
corresponding to $p_\text{max}$ is set to one, and (2) an independent random
number $r$, extracted with uniform probability in $[0,1)$, turns out to be
smaller than $p_\text{min}/p_\text{max}$.

Finally, we need to discuss our computation of the random integers $k$. In
fact, we need to generate 256 independent $k$ numbers, because we codify 256
spins in our superspins. After some reflection, we decided to use the {\tt
  xoshiro256++} generator~\cite{blackman:19}, which ensures the same level of
randomness on each of its 64-bits. We employed a 256-bits streaming
extension to implement a parallel version of {\tt xoshiro256++}, composed of
four independent {\tt xoshiro256++} random sequences.  Hence, each call to our
generator produces 256 independent random bits which are 1 with $50\%$
probability. Therefore, $m$ calls to {\tt xoshiro256++} produces 256
independent $k$ random numbers.

In our simulations, the optimal value of $m$ has turned out to be $m\simeq
10$. In this way, we have found it possible to compute the spin-flip bit for
1024 spins with approximately 42 calls to our random number generators (namely,
40 calls for $d_1$ and two calls for the computation of $d_2$).

%
%
\section{\texorpdfstring{Space integrals of $\bm{C(\bm{r};t)}$ correlation}%
                        {Space integrals of C(r,t) correlation}}
\label{Sec:integrals}

At long distances, the correlation function $C(\bm{r};t)$ decays as
$f(-r/\xi(t))/r^a$, where $\xi$ is the coherence length, $f(x)$ is a cut-off
function, and $a$ is an exponent. Out-of-equilibrium, the function $f(x)$
decays super-exponentially, while, at equilibrium, the decay is only
exponential. Unfortunately, the exponent $a$ and the precise form of the
function $f(x)$ are known only at equilibrium. Nevertheless, following
Ref.~\cite{janus:08b}, we can obtain $\xi(t)$ from an integral estimator.

Hereafter, we shall use $C(r;t)$ as a shorthand for $C(\bm{r};t)$ with
$\bm{r}=(r,0)$ or $\bm{r}=(0,r)$. Introducing the integrals
\begin{equation}\label{eq:int}
I_k(t)\;=\;\int_0^\infty r^kC(r;t)\, \mathrm{d}r\, ,
\end{equation}
we can define $\xi_{k,k+1}(t)\equiv I_{k+1}(t)/I_k(t)$, which is proportional
to $\xi(t)$.  In our case, we have chosen as an estimator of $\xi(t)$ the
ratio $\xi_{1,2}(t)$, which has been well studied in the literature (see e.g.,
Refs.~\cite{fernandez:18b,janus:18,fernandez:19}, and references therein).
 
The difficulty in the computation of this integrals arises from the big
relative errors $[\Delta C(r;t)/C(r;t)]$ present in the large-$r$ tail of
$C(r;t)$ (this problem is well known in the context of the analysis of
autocorrelation times in MC simulations~\cite{sokal:97}). Our
solution, which is inspired by
Ref.~\cite{fernandez:18b,janus:18,fernandez:19}, combines two strategies: (1)
reduction of the relative error in the correlation function by considering a
sufficiently large number of replicas (see Sec.~\ref{sec:thermal}), and
(2) the use of a smart way to estimate $I_k(t)$. Our estimation of $I_k(t)$ is
the sum of two contributions: the numerical integral of our measured $C(r;t)$
up to a noise-dependent cut-off, and a tail contribution estimated by using a
smooth extrapolation function, namely, $F(r)=A\text{e}^{-(r/\xi_F)^\beta}/r^b$
with $b=1/2$.

Strictly speaking, the functional form of $F(r)$ is only valid in the
paramagnetic phase~\cite{fernandez:18b}.  In the ferromagnetic phase, the
exponent of the algebraic term $r^b$ is not known.  However, because the
exponential term decays very fast ($\beta\approx 2$), the precise value of $b$
is not very relevant.  On the other hand, when the equilibrium is reached in
the paramagnetic phase, where $\beta=1$, the slow decay of $C(r,t)$ makes the
tail contribution very relevant (this is why we have needed a large number of
replicas in order to have a good estimation of the tail contribution in 
the paramagnetic phase).

The complete procedure to compute $I_k$ is as follows:
\begin{enumerate}
\item We obtain a spatial cut-off determined by the size of our statistical
  errors. The noise-dependent distance $r_\mathrm{cut}$ is the shortest
  distance such that $C(r_\mathrm{cut}+1;t)$ is smaller than three times its
  error. Of course, $r_\mathrm{cut}$ is time-dependent.

\item Estimate the region $[r_\mathrm{min},r_\mathrm{max}]$ employed in the
  fit to $F(r)$.  We start by locating the position $r^*$ of the maximum of
  the quantity $r^2C(r;t)$; the value of such maximum is
  $\rho={r^*}^2C(r^*;t)$.  Next, we determine $r_\mathrm{min}$ and
  $r_\mathrm{max}$, which must verify the inequalities
  $r^*<r_\mathrm{min}<r_\mathrm{max}<r_\mathrm{cut}$.  We define
  $r_\mathrm{min}$ (resp.\/ $r_\mathrm{max}$) as the first $r$ where
  $r^2C(r;t)<a_\mathrm{min}\rho$ (resp.\/ $r^2C(r;t)<a_\mathrm{max}\rho$).  We
  take $a_\mathrm{min}=0.9$ and $a_\mathrm{max}=1/3$. We regard a failure to
  meet the condition $ r_\mathrm{max}<r_\mathrm{cut}$ as a sure-tell sign of
  the need of increasing the number of replicas.
\item Finally we calculate the integrals. There are two possibilities (in both
  cases we estimate errors with the jackknife
  method~\cite{amit:05,yllanes:11}):
  \begin{enumerate}
   \item If $r_\mathrm{max}-r_\mathrm{min}>8$, we fit $C(r;t)$ in the interval
     $[r_\mathrm{min},r_\mathrm{max}]$ to $F(r)$ (the fit parameters are the
     amplitude $A$, the length scale $\xi_F$, and the exponent $\beta$). We
     estimate the integral as the sum of two contributions, namely, the
     integral of $C(r;t)$ from $r=0$ to $r_\mathrm{max}$ and the integral of
     $F(r)$ from $r_\mathrm{max}$ to $r=20\xi_F$. In both terms, we use a
     second-order Gaussian quadrature (when an interpolation of $C(r;t)$ is
     needed, we employ a cubic spline interpolation).
   \item If the region $[r_\mathrm{min},r_\mathrm{max}]$ is small
     (i.e. $r_{\mathrm{max}}-r_{\mathrm{min}}\leq 8$), we integrate only
     numerically $C(r;t)$ up to $ r_\mathrm{cut}$.
  \end{enumerate}
\end{enumerate}

\input{Ising2D_regular_v1.bbl}

\end{document}

%% file: Ising2D_regular_v1.bbl
%